\documentclass[12pt]{article}
\usepackage{amsmath}
\usepackage{amsthm}
\usepackage{amssymb}
\usepackage{graphicx}
\usepackage{setspace}
\usepackage{verbatim}
\usepackage{subfiles}
\usepackage{tikz}
\usepackage[disable]{todonotes}
\usepackage{enumerate}
\usepackage{subcaption}	
\allowdisplaybreaks
\newtheorem{theorem}{Theorem}
\newtheorem{lemma}[theorem]{Lemma}

\newcommand{\sX}{\mathsf{X}}

\theoremstyle{remark}

\usepackage[sort,longnamesfirst]{natbib}
\setcitestyle{authoryear,open={((},close={))}}
\usepackage{geometry} 
\newcommand {\real} {\mathbb{R}}

\def\baro{\vskip  .2truecm\hfill \hrule height.5pt \vskip  .2truecm}

\title{Directional Metropolis-Hastings}
\author{Abhirup Mallik \\School of Statistics \\
  University of Minnesota \\
  {\tt malli066@umn.edu} \and Galin L. Jones \footnote{Research supported
    by the National
    Institutes of Health and the National Science Foundation.}\\
  School of Statistics \\
  University of Minnesota \\
  {\tt galin@umn.edu} } \date{\today}

\graphicspath{{images/}{../images/}}
\begin{document}

\maketitle

\begin{abstract}
	We propose a new kernel for Metropolis Hastings called Directional Metropolis Hastings (DMH) with multivariate update where the proposal kernel has state dependent covariance matrix. We use the derivative of the target distribution at the current state to change the orientation of the proposal distribution, therefore producing a more plausible proposal. We study the conditions for geometric ergodicity of our algorithm and provide necessary and sufficient conditions for convergence. We also suggest a scheme for adaptively update the variance parameter and study the conditions of ergodicity of the adaptive algorithm. We demonstrate the performance of our algorithm in a Bayesian generalized linear model problem.
\end{abstract}

\newpage

\section{Introduction} 
\label{sec:intro}
Markov Chain Monte Carlo (MCMC) is a standard method for generating instances of random variables from a probability distribution or estimating expectation of a functional with respect to a probability distribution. A common algorithm to perform MCMC is using the Metropolis Hastings Algorithm. In this paper we propose a new variant of Metropolis Hastings algorithm that generalizes the standard random walk Metropolis Hastings and offers a greater level of flexibility to the practitioner to tune the algorithm for producing an optimal chain especially for multivariate target distributions. The generalization comes in the form of a multivariate normal proposal kernel for Metropolis Hastings with covariance matrix dependent on the state. We describe the motivation and the construction of this new kernel and study the stability properties of the resulting algorithm.

One common problem for statisticians and various other branches of science is finding expectation of a functional with respect to a probability density. Let $f$ be a positive probability density function on $\sX \subseteq \real^{d}$, $d
\ge 1$.  When $f$ is intractable in the sense that expectations 
\[
\mu := \int_{\sX} g(x) f(x) \, dx 
\]
are difficult to calculate, it is commonplace to turn to Markov chain
Monte Carlo (MCMC) methods.  The fundamental MCMC  algorithm is 
Metropolis-Hastings, \citet{hastings} which is now described.  Let $p(\cdot, \cdot)$
denote the proposal density.  If $X_{n}=x$ denotes the
current state of the simulation, then the next state is obtained as
follows.

\baro
Iteration $n+1$ of Metropolis-Hastings

\begin{enumerate}
\item Draw $Y \sim p(x, \cdot)$ and independently $U \sim
  \text{Uniform}(0,1)$.  Call the observed values $y,u$.
\item If 
\[
u \le \frac{f(y) p(y, x)}{f(x) p(x,y)}
\]
set $X_{n+1}=y$.  Otherwise, set $X_{n+1}=x$.
\end{enumerate}
\baro

The effectiveness of Metropolis-Hastings is controlled by the choice
of proposal distribution.  The most common ways to choose proposals do
not take the structure of $f$ into account.  For example, if $p(y)$
does not depend on the current state, then an \textit{independence}
Metropolis-Hastings sampler results while if $p(x,y)=p(y,x)$, then we
have a \textit{symmetric} Metropolis-Hastings sampler.  If
$p(x-y)=p(y-x)$, then we have a \textit{random walk}
Metropolis-Hastings sampler, denoted RWMH.  A particularly popular
choice of random walk sampler uses a multivariate Normal density
centered at the current state, that is, $\text{N}_{d}(x,
\sigma^{2}I_{d})$ where $\sigma^{2}$ is a tuning parameter. The
Metropolis-adjusted Langevin algorithm (MALA) \citet{roberts1996} uses a $\text{N}_{d}(x +
h \nabla \log f(x), h^{2}I_{d})$ proposal density with $h$ a tuning
parameter which is typically small.

While any of the common choices of proposal can work well, their use
will often produce a slowly mixing chain so that enormous simulation
sizes are required to produce a small effective sample size.  We
introduce a method for choosing a proposal distribution that exploits
information available about the target density.  More specifically, we
use a Normal density whose mean is a function of the gradient of $f$
and a covariance matrix constructed to mimic the shape of $f$. As a notational clarification, $||.||$ means Euclidean norm.  Let 
\[
g(x)=\frac{\nabla \log f(x)}{\|\nabla \log f(x)\|} \; .
\]
and $G(x)=g(x) g(x)^{T}$. Then the proposal distribution we
consider is 
\[
\text{N}_{d}(x + h \nabla \log f(x), tI_{d} + (s-1)
G(x))
\]
where $t > 0$, $h \geq 0$ and $s > 0$ are tuning parameters.  The construction of
$\Sigma(x) = tI_{d} + (s-1) G(x)$ will be fully described and
justified in Section~\ref{sec:dmh}.  We call the resulting algorithm
\textit{directional Metropolis-Hastings} (DMH).

DMH reduces to RWMH for $h=0,s=1$, and it becomes a special case of MALA for $s=1$.

Due to the need to construct $\Sigma_{x}$ at each step DMH certainly
requires more computational resources than RWMH, but it is comparable
to MALA.  Moreover, we will see in several examples that it enjoys
better empirical performance than both RWMH and MALA.    

\begin{figure}
	\includegraphics{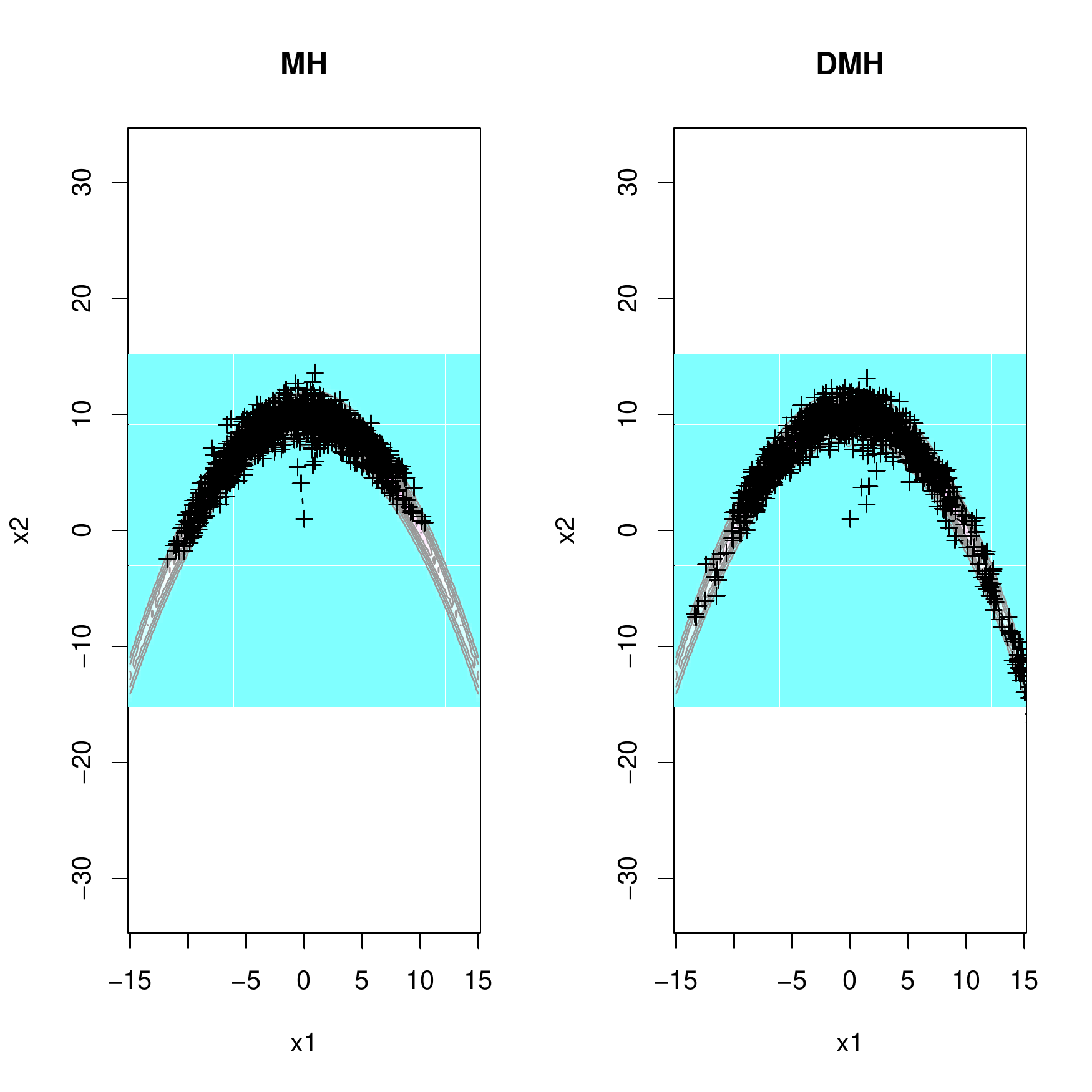}
	\caption{Plot of the generated chain for the Banana shaped distribution.}
	\label{banana}
\end{figure}

To visualize the behavior of this kernel, we use it on a distribution with a irregular shaped contours. The following distribution is proposed by \citet{Haario:2001,Haario:1999} with density
\begin{align*}
f_B(x_1, \cdots, x_d) \propto \exp[-x_1^2/200 - \frac{1}{2}(x_2 + Bx_1^2 - 100B)^2 - \frac{1}{2}(x_3^2 + \cdots + x_d^2) ]
\end{align*}

with $B > 0$, can be called as the \emph{Bananacity} constant. In figure \ref{banana} we show the generated chains of length $2000$ on the contours of the target density in two dimensions and compare with RWMH and DMH kernel. It seems that the DMH algorithm is better in exploring the full state space than RWMH. Because of the directional nature of the proposal density, we would expect that the algorithm is much less likely to get stuck at some point and it should be able to sample from the tail areas much better. 

In the reminder of the paper, we discuss in detail about the construction of the DMH kernel, followed by a study of it's convergence properties. We then discuss the examples where this kernel performs better and specifically demonstrate it's performance in ridge penalized generalized linear models estimation problems. We also propose an adaptive scheme to update the scale parameter and discuss the conditions of ergodicity in the adaptive algorithm.

\section{Construction of DMH Kernel} 
\label{sec:dmh}
Here we describe the motivation behind constructing the DMH kernel. Let $x\in \mathbb{R}^d$ be the current state of the Markov chain. We need the following elements to construct the adaptive proposal at $x$. We define a re-weighted norm for our algorithm as:

For a given $0 < s $,
\begin{align*}
S = diag\{1/s,1,\cdots,1\}\\
||x||_s^2 := x^T S x
\end{align*}

The above norm is just a weighted sum of squares. Notice that, for $s=1$, this norm reduces to Euclidean norm. We have defined $g(x)$ as the scaled gradient of the log unnormalized target distribution with unit euclidean norm. We now need to find the orthogonal complements of $g(x)$. Let $\{g_1(x), \cdots, g_{p-1}(x)\}$ be the completion of the vector $g(x)$, so that $\text{span}(g(x),g_1(x), \cdots, g_{p-1}(x)) = \mathbb{R}^d$. We define the matrix $G_d(x)$ with columns of $(g(x), g_1(x), \cdots, g_{d-1}(x))$.

One way to implement this completion of basis will be by using Gram–Schmidt orthogonalization process. We can think of $G(x)$ as a basis for $\mathbb{R}^d$, with the first vector directed toward the gradient of the log unnormalized target density. We want to put different relative importance to that direction while constructing the proposal covariance, hence we use a weighted combination of $G_d(x)$ as follows:
\begin{align*}
\Lambda = diag\{s,1,\cdots,1\}\\
\Sigma(x) = G_d(x)\Lambda G_d(x)^T
\end{align*}
We want to construct a normal distribution centered at: $x+C(x)$, where $C(x) = h \nabla_x \log \pi (x)$ and with covariance matrix $\Sigma(x)$. 
We show the acceptance ratio for the ease of implementation. Notice that, if the proposed value is $y$, then, $|\Sigma_x| = |\Sigma_y| = s$, where $|.|$ indicates the value of the determinant.
\begin{align*}
\alpha(x,y) =& \frac{\pi(y)}{\pi(x)} \exp\{-\frac{1}{2}[||G_y(x-(y+C(y)))||_s^2-||G_x(y-(x+C(x)))||_s^2]\}\\
\end{align*}

We first provide a lemma that ensures the uniqueness of $\Sigma(x)$ constructed at each state from the gradient. It is constructed using a basis completion method, and if it was not unique, then Metropolis Hastings algorithm might not have been reversible. Even though $\{g_1(x), \cdots, g_{d-1}(x)\}$ is not unique, producing a $G_d(x)$ that is also not unique, however the proposal kernel constructed in this method is unique, as shown in the following lemma.

\begin{lemma}\label{lem:sigma}
If $\Sigma(x) = G_d(x)\Lambda G_d(x)^T$ is constructed as mentioned above and $\Lambda = \text{diag}\{s,1, \cdots, 1\}$. Then, 
\begin{align*}
\Sigma_x = I_d + (s-1)g(x)g(x)^T
\end{align*}
\end{lemma}

\begin{proof}
See Appendix~\ref{app:sigma lemma}.
\end{proof}

\section{Stability Properties of DMH} 
\label{sec:stability_properties_of_dmh}
Once we have developed the MCMC algorithm, we need to ensure that the chain produced converges to the right target distribution. To set up notation, we will work with the measurable space $(\mathcal{X},\mathcal{B})$ for a discrete-time Markov chain $\{X_t\}_{t\geq 0}$. The time homogeneous transition kernel is $P:\mathcal{X} \times \mathcal{B} \to [0,1]$, with $n$ step transition probability defined as 
\begin{align*}
P^n(x,A) = P[X_n \in A | X_0 = x]
\end{align*}
If the MCMC algorithm is doing the right thing, then we should expect $P^n(x,A)$ to be close to $f(A)$ as $n$ becomes large enough, with $f(.)$ being the target density. The closeness in probability is measured in \emph{Total Variation} norm, where
\begin{align*}
||\mu(.) - \nu(.)||_{TV} = \sup_{A \in \mathbb{B}} |\mu(A) - \nu(A)|
\end{align*}
The transition probability in our algorithm is controlled by hastings ratio and under the standard conditions of $\phi$-irreducible and aperiodicity, using results from Chapter 13 of \citet{meyn:twee:2009}, for almost all initial starts $x_0 \in \mathcal{X}$,
\begin{align*}
\lim_{n \to \infty} ||P^n(x,.) - f(.)||_{TV} = 0
\end{align*}

Even though the convergence to the target distribution is ensured in an MCMC algorithm, the speed of convergence is also of interest. A Markov chain is called Geometrically Ergodic if for $f$-almost all $x \in \mathcal{X}$

\begin{align*}
||P^n(x,.) - f(.)||_{TV} \leq M(x)\gamma(n)
\end{align*}
where, $M(x)$ is a nonnegative function and $\gamma(n)$ is a non negative decreasing function of $n$. If a Markov chain is Geometrically Erogodic, and if $E_f[|g|^{2+\delta}]$, then we have the Markov chain central limit theorem:

\begin{align*}
\sqrt{n}(\bar{g}_n - E_{f}g) \xrightarrow{\mathcal{D}} N(0,\Sigma_{f,g})
\end{align*}
The asymptotic variance term $\Sigma_{f,g} \text{Var}_{f}\{g(X)\} + 2\sum_{i=1}^{\infty} \text{Cov}_{f}\{g(X_0),g(X_i)\}$. The existence of CLT allows us to use the asymptotic variance to assess the quality of the estimate. Details about estimating the asymptotic variance is discussed in \citet{Vats2015}. So establishing conditions for Geometric Ergodicity is a crucial component of a MCMC algorithm.

To show the Geometric Ergodicity of our algorithm we follow similar lines of argument from  \citet{robe:twee:1996}. This approach relies on finding an appropriate drift function and a small set to satisfy the Geometric Drift Condition as mentioned in Chapter 15 of \citet{meyn:twee:2009}. For ease of computation, we divide our state space into acceptance and rejection regions in the lines of  \citet{meng:twee:1996}. We write the acceptance region of MH from the point $x$ as $A(x)$, that is $A(x)$ is the region where the proposals are always accepted. Hence,
\begin{align*}
A(x) = \{y: \alpha(x,y) = 1\}
\end{align*}
We define, $R(x) = A(x)^c$ as the possible rejection region. And we
borrow the concept of Inward Convergence from
\citet{roberts1996}. The Interior of a point $x$ is defined as:
\begin{align*}
I(x) = \{y: ||y|| \leq ||x||\}
\end{align*}
It is said that $A(.)$ converges inward in $P$, if,
\begin{align*}
\lim_{||x|| \to \infty} \int_{A(x)\Delta I(x)} P(x,y)dy = 0
\end{align*}

Here the symmetric difference set operator is denoted by $A\Delta B = (A  \cup B) \setminus (A \cap B)$
\begin{theorem} \label{thm:ge}
Let $\mu(x) = x + h \nabla \log f(x)$ be the 'center' of the candidate density. Let us define:
\begin{align*}
\eta \equiv \liminf_{|x| \to\infty} (||x|| - ||\mu(x)||)
\end{align*}
If it is assumed that $A(.)$ converges inwards in $q$, then for $V_{\tau}(x) = e^{\tau||x||}$, and for $0 < s \leq 1$ directional MH is $V_{\tau}$-uniformly ergodic for $\tau < h\eta$
\end{theorem}

\begin{proof}
See Appendix~\ref{app:ge thm}.
\end{proof}
\todo{Explain more about terms}

The details about inward convergence property and how it can possibly be relaxed is given in  \citet{roberts1996}. In fact the condition that $\eta > 0$ can also be ensured by imposing some conditions on $\nabla f(x)$. 
Even though the above result provides sufficient condition for Geometric Ergodicity of our algorithm, it is not always easy to check the conditions needed. In general it is difficult to verify the conditions for geometric ergodicity for any MCMC algorithm. So, in this section, we try to provide some conditions that are necessary for establishing geometric ergodicity. In a way these conditions give us a way for quickly verifying the lack of geometric ergodicity. 

The result is based on the idea of "random walk type" Markov Chains. We say that $X$ is of random walk type if, for every $\epsilon > 0$, there exists $K > 0$, such that $P(x,B(x,K) > 1 - \epsilon$ for all $x$, where $B(x,K) = \{y: |y - x| < K\}$ denotes the open ball with radius $K$ and center at $x$. Chains of random walk type are very common in MCMC and this idea is needed to prove the necessary conditions mentioned in the following theorem.

\begin{theorem}
	Assume that $f(.)$ is a strictly positive and twice differentiable density. Then if DMH produces a geometrically ergodic chain $X$, then there exists $s > 0$ such that,
	\begin{align*}
	\int_{\mathbb{R}^d} \exp\{s||x||\} f(\text{dx}) <& \infty
	\end{align*}
	\label{thm:rwtype}
\end{theorem}

\begin{proof}
Appendix C
\end{proof}

\section{Examples} 
\label{sec:examples}
We have implemented DMH algorithm in an R package called \emph{dirmcmc} \citet{dirmcmc} available via CRAN. Here we demonstrate its performance in a Bayesian regression  problem. We consider one parameter exponential family as follows. Let $y_i$'s be independent random variables with
\begin{eqnarray}
f(y_i|\theta_i) &=& \exp[a^{-1}(\phi_i)\{y_i\theta_i - \psi(\theta_i)\} + c(y_i;\phi_i)]
\end{eqnarray}
We assume that $\theta_i = h(x_i^T\beta + u)$, where $h$ is a sufficiently smooth function, which in our example is just the identity function. The $\psi(.)$ function changes based on the link function for the GLM and we show all different combinations of $\psi(.), a, h(.)$ in table \ref{modeltab} Let $X^T = (x_1, \cdots, x_n)$, be a $n \times p$ matrix, where each predictor is a $p$ dimensional vector. The log likelihood is given by:
\begin{eqnarray}
	\log L (\beta,u) &=& \sum_{i = 1}^n a^{-1}(\phi_i)\{y_ih(x_i^T\beta + u) - \psi(h(x_i^T\beta + u))\}\\
	\nabla \log L(\beta,u) &=& \sum_{i = 1}^n a^{-1}(\phi_i) \{y_i - \psi'(h(x_i^T\beta + u))\}
\end{eqnarray}
We use the following priors
\begin{eqnarray}
(\beta|u,v_u) &\sim& N(0,v_BI_p)	\\
(u|v_u) &\sim& N(0,v_u)
\end{eqnarray}
The log posterior can be written as:
\begin{align*}
\log (\pi(\beta,u,v_u|Y=y)) = & \sum_{i = 1}^n [a^{-1}(\phi_i)\{y_ih(x_i^T \beta + u_i) - \psi(h(x_i^T \beta + u_i))\}\\
 &- \{ (\frac{ \beta^T \beta}{2v_{ \beta}} + \frac{u^Tu}{2v_u} \}]
\end{align*}

\begin{table}[ht]
	\caption{Three common members of exponential family regression}
	\label{tab:exptable}
	\begin{center}
		\begin{tabular}{l|c|c|c}
		\hline

		\hline
		\textbf{Family} & \textbf{$h(x_i^T \beta + u)$} & \textbf{$a(\phi)$} & \textbf{$\psi(h(x_i^T \beta + u))$} \\
		\hline
			Normal & $x_i^T \beta + u$ & $\sigma^2$ & $\frac{1}{2}(x_i^T \beta + u)^2$ \\
		\hline
			Bernoulli & $x_i^T \beta + u$ & $1$ & $\log[1 + \exp(x_i^T \beta + u)]$ \\
		\hline
			Poisson & $x_i^T \beta + u$ & $1$ & $\exp(x_i^T \beta + u)$ \\
		\hline
		\end{tabular}
	\end{center}
  \label{modeltab}
\end{table}
We can calculate the derivatives of the log posterior with respect to our parameters as:
\begin{align*}
\nabla_{\beta} \log(\pi(\beta,u,v_u|Y=y)) &= [\sum_{i = 1}^n a_i^{-1}(\phi_i) \{y_i - \nabla_{\beta} \psi(h(x_i^T \beta + u))\} \nabla_{\beta}h(x_i^T + u)x_i - \beta/v_{\beta}]\\
\nabla_{u} \log(\pi(\beta,u,v_u|Y=y)) &= [\sum_{i = 1}^n a_i^{-1}(\phi_i) \{y_i - \nabla_{u} \psi(h(x_i^T \beta + u))\}\nabla_{u}h(x_i^T + u) - u/v_u]
\end{align*}

In our simulated experiment, we generated the predictors independently from standard normal distribution, the true values of the coefficients were generated from uniform distribution, and independent standard normal noise was added to generate the responses. We chose a sample size of $100$. The chains were run for a length of $10^5$. The scaling parameter for RWMH and DMH are kept the same to keep the comparisons sensible.

We compare algorithms based on the quality of the chain they have produced. Even though there is no consensus in the field of MCMC about any single metric describing the "quality" of the resulting chain, there are several metrics that are available in the literature and are fairly standard to use. One of the measure of quality of mixing for Markov chains is Integrated Autocorrelation times, relative to i.i.d. samples, given by
\begin{align*}
\text{ACT}_{\text{est}} =& 1 + 2 \sum_{i=1}^l \hat{\gamma}_i
\end{align*}
Where $\hat{\gamma}_i$ is the estimated autocorrelation of lag $i$. The sum is truncated at step $l$, where the autocorrelation drops below $0.05$.

Effective sample size is another measure commonly used in this context. Coordinate wise ESS is defined as for $i$th coordinate,
\begin{align*}
\hat{ESS}_i =& n \frac{\lambda_{n,i}^2}{\sigma^2_{n,i}}
\end{align*}
Where $\sigma^2_{n,i}$ is a strongly consistent estimator of variance of $i$th component, and $\lambda_{n,i}^2$ is the sample variance of the $i$th component of the chain. \citet{Vats2015} defines a multivariate analog of the above definition as
\begin{align*}
\hat{mESS} =& n \Big( \frac{|\nabla_n|}{|\Sigma_n|} \Big)^{1/p}
\end{align*}
For a detailed description of properties of ESS and mESS we refer \citet{Vats2015}.

An alternative efficiency measure for multivariate chains is given by the Mean Squared Jumping Distance (MSJD), which is defined by
\begin{align*}
\text{MSJD} =& \frac{1}{(n-1)}\sum_{i=1}^{n-1}||X^{(i+1)} - X^{(i)}||_2^2
\end{align*}
Here we are considering Euclidean norm, and the expectation of this quantity is called Expected square Euclidean Jump Distance. For a stationary chain, maximizing ESEJD is equivalent to minimizing a weighted sum of the lag-1 autocorrelations.

\todo{Interpret the results}
In table \ref{tab:bayesglmtab2} we compare their performance as a multivariate chain. All the results mentioned here are from chain of running length $10000$. In table \ref{tab:esscomp} we compare the component wise effective sample size and in table \ref{tab:iactcomp} we compare integrated auto correlation times per chain. We can clearly see the improvements in MultiESS for all the models, and this is also reflected in \ref{tab:esscomp} for component wise effective sample size gains. IACT have clearly reduced and we see some marginal gains in MSJD. 

\begin{table}[ht]
\centering
\begin{tabular}{|c|c|c|c|c|}
  \hline
 Algorithm & Model & Acceptance & MultESS & MSJD \\ 
  \hline
Normal & RWMH & 0.31 & 566.66 & 0.58 \\ 
 Normal & DMH & 0.51 & 4304.01 & 2.00 \\ 
 Normal & ADMH & 0.45 & 3290.06 & 2.29 \\ 
  \hline
 Bernoulli & RWMH & 0.35 & 511.08 & 118.78 \\ 
 Bernoulli & DMH & 0.36 & 1735.07 & 395.93 \\ 
  Bernoulli & ADMH & 0.42 & 2550.72 & 336.59 \\ 
  \hline
  Poisson & RWMH & 0.15 & 348.63 & 0.01 \\ 
  Poisson & DMH & 0.27 & 339.61 & 0.01 \\ 
  Poisson & ADMH & 0.40 & 506.68 & 0.00 \\ 
   \hline
\end{tabular}
\caption{Comparison of RWMH, DMH and ADMH for Bayesian GLM model.} 
\label{tab:bayesglmtab2}
\end{table}

\begin{table}[ht]
\centering
\begin{tabular}{|l|l|r|r|r|}
  \hline
 Algorithm & Variable & Normal & Bernoulli & Poisson \\ 
  \hline
   RWMH & $\beta_1$ & 537.602 & 405.852 & 218.143 \\ 
    DMH & $\beta_1$ & 3774.931 & 950.676 & 206.756 \\ 
   ADMH & $\beta_1$ & 4178.928 & 1133.967 & 337.313 \\ 
   \hline
   RWMH & $\beta_2$ & 535.592 & 399.498 & 172.253 \\ 
    DMH & $\beta_2$ & 3850.913 & 783.425 & 191.859 \\ 
   ADMH & $\beta_2$ & 2866.576 & 1012.666 & 259.964 \\ 
   \hline
   RWMH & $\beta_3$ & 179.827 & 358.477 & 178.023 \\ 
    DMH & $\beta_3$ & 3073.830 & 1335.223 & 281.655 \\ 
   ADMH & $\beta_3$ & 2087.911 & 1388.001 & 383.036 \\ 
   \hline
   RWMH & $\beta_4$ & 535.936 & 568.968 & 115.888 \\ 
    DMH & $\beta_4$ & 3013.423 & 2395.125 & 239.805 \\ 
   ADMH & $\beta_4$ & 3033.364 & 4149.616 & 314.272 \\ 
   \hline
   RWMH & $\beta_5$ & 314.900 & 499.739 & 125.298 \\ 
    DMH & $\beta_5$ & 4274.138 & 1340.612 & 144.406 \\ 
   ADMH & $\beta_5$ & 3379.095 & 2099.024 & 199.748 \\ 
   \hline
\end{tabular}
\caption{Comparison of Effective sample size of each component for three MCMC algorithms.}
\label{tab:esscomp}
\end{table}

\begin{table}[ht]
\centering
\begin{tabular}{|l|l|r|r|r|}
  \hline
  Algorithm & Variable & Normal & Bernoulli & Poisson \\ 
  \hline
   RWMH & $\beta_1$ & 23.752 & 30.091 & 76.255 \\ 
    DMH & $\beta_1$ & 3.878 & 13.049 & 91.293 \\ 
   ADMH & $\beta_1$ & 4.314 & 9.540 & 37.353 \\ 
   \hline
   RWMH & $\beta_2$ & 25.530 & 33.591 & 68.275 \\ 
    DMH & $\beta_2$ & 4.297 & 15.771 & 93.646 \\ 
   ADMH & $\beta_2$ & 4.646 & 11.077 & 51.661 \\ 
   \hline
   RWMH & $\beta_3$ & 60.136 & 32.715 & 71.516 \\ 
    DMH & $\beta_3$ & 5.126 & 10.603 & 40.440 \\ 
   ADMH & $\beta_3$ & 5.632 & 7.921 & 28.724 \\ 
   \hline
   RWMH & $\beta_4$ & 23.385 & 19.731 & 132.969 \\ 
    DMH & $\beta_4$ & 4.759 & 6.614 & 60.785 \\ 
   ADMH & $\beta_4$ & 5.311 & 4.827 & 42.224 \\ 
   \hline
   RWMH & $\beta_5$ & 53.116 & 23.691 & 221.451 \\ 
    DMH & $\beta_5$ & 4.313 & 9.404 & 196.012 \\ 
   ADMH & $\beta_5$ & 4.885 & 6.063 & 73.342 \\ 
   \hline
\end{tabular}
\caption{Comparison of Integrated Auto Correlation times of each component for three MCMC algorithms.}
\label{tab:iactcomp}
\end{table}


\section{An extension to Adaptive MCMC} 
\label{sec:adaptive}
Any MCMC algorithm requires fine tuning of the parameters associated with the proposal kernel. Adaptive MCMC proposes to solve this problem by automatically updating the tuning parameters to reach some target metric of performance. The major tuning parameter for our algorithm is the multiplier to the proposal variance for our kernel. Although, there are other parameters in our algorithm, we found updating this parameter adaptively resulted in improvement of the resulting chain.

There are several schemes of adaptations available in the literature, however only a few of them are applicable for multivariate updates. We refer to \citet{roberts:09} for a survey of various adaptive schemes in MCMC. Adaptive MCMC have their own limitations as well, as it is known that adaptive MCMC may not preserve stationarity of target distribution. However, \citet{roberts2007} have proposed two conditions which are simpler to apply and does not require the adaptive parameters to converge themselves. We summarize the two conditions here.

Let $\{P_{\gamma}\}_{\gamma \in \mathcal{Y}}$ be a collection of Markov chain kernels on $\mathcal{X}$, each of which has stationary distribution as the target distribution. If the algorithm updates $X_n$ to $X_{n+1}$ using the kernel $P_{\Gamma_n}$, then we refer to $\mathcal{Y}$ as the adaptation index. By theorem 1 and corollary 5 of \citet{roberts2007} it is guaranteed that the adaptive algorithm would be ergodic if 
\begin{itemize}
	\item (\textit{Diminishing Adaptation}) 
	\begin{align*}
		\lim_{n \to \infty} \sup_{x \in \mathcal{X}} ||P_{\Gamma_{n+1}}(x,.) - P_{\Gamma_n}(x,.)|| &= 0 
	\end{align*}
	\item (\textit{Bounded Convergence}) Let $\epsilon$ convergence time function $M_{\epsilon}:\mathcal{X} \times \mathcal{Y} \to N, \epsilon >0 $ by 
	\begin{align*}
		M_{\epsilon}(x,\gamma) &= \inf \{n \geq 1: ||P_{\gamma}^n(x,.) - \pi(.)|| \leq \epsilon\} 
	\end{align*}
	Then \textit{(Bounded Convergence)} is that the sequence $\{M_{\epsilon}(x,\gamma)\}_{n=0}^{\infty}$ is bounded in probability.
	
\end{itemize}

\citet{roberts2007} have shown that \textit{(Bounded Convergence)} is satisfied whenever $\mathcal{X}\times\mathcal{Y}$ is finite, or is compact in some space in which either the transition kernels $P_{\gamma}$, or the Metropolis Hastings proposal kernels $Q_{\gamma}$, have jointly continuous densities. 

\subsection{Adaptive scheme} 
\label{sub:adaptive_scheme}
We use a batchwise adaptation strategy. We try to achieve a target acceptance probability, which can also be specified by the user. In some problems, optimal acceptance probability is theoretically know, while there are also recommended ranges of acceptance that the MCMC practitioners are aware of. The algorithm proceeds as follows:
\baro
Given a batch size $B$, target acceptance rate $a$, for the batch $b$ the scale parameter $\sigma$ is adapted as follows:
\begin{align*}
\log(\sigma_{b+1}) \leftarrow   \begin{cases} 
   \log(\sigma_b) + \delta(b) & \text{if acceptance rate} \geq a  \\
   \log(\sigma_b) - \delta(b) & \text{if acceptance rate} < a 
  \end{cases}
\end{align*}
We choose the update function as $\delta(b) = \min(0.01, b^{-1/2})$. $\{\log(\sigma_b)\}$ is restricted in $[-M,M]$
\baro
\subsection{Ergodicity of Adaptive Algorithm} 
\label{sub:ergodicity_of_adaptive_algorithm}
We use the conditions given above to check how likely it is that the adaptive scheme preserves the ergodicity. Because this scheme of adaption is already proposed for coordinate wise update in \citet{roberts2007}, we follow a similar reasoning. In the lemma below we give the conditions for verifying the two conditions.
\begin{lemma}
	The adaptive MCMC algorithm with update function $\delta(b) = \min(0.01, b^{-1/2})$ is ergodic if the target density is log concave with bounded support.
\end{lemma}

\begin{figure}
\begin{subfigure}{.5\textwidth}
  \centering
  \includegraphics[width=\linewidth]{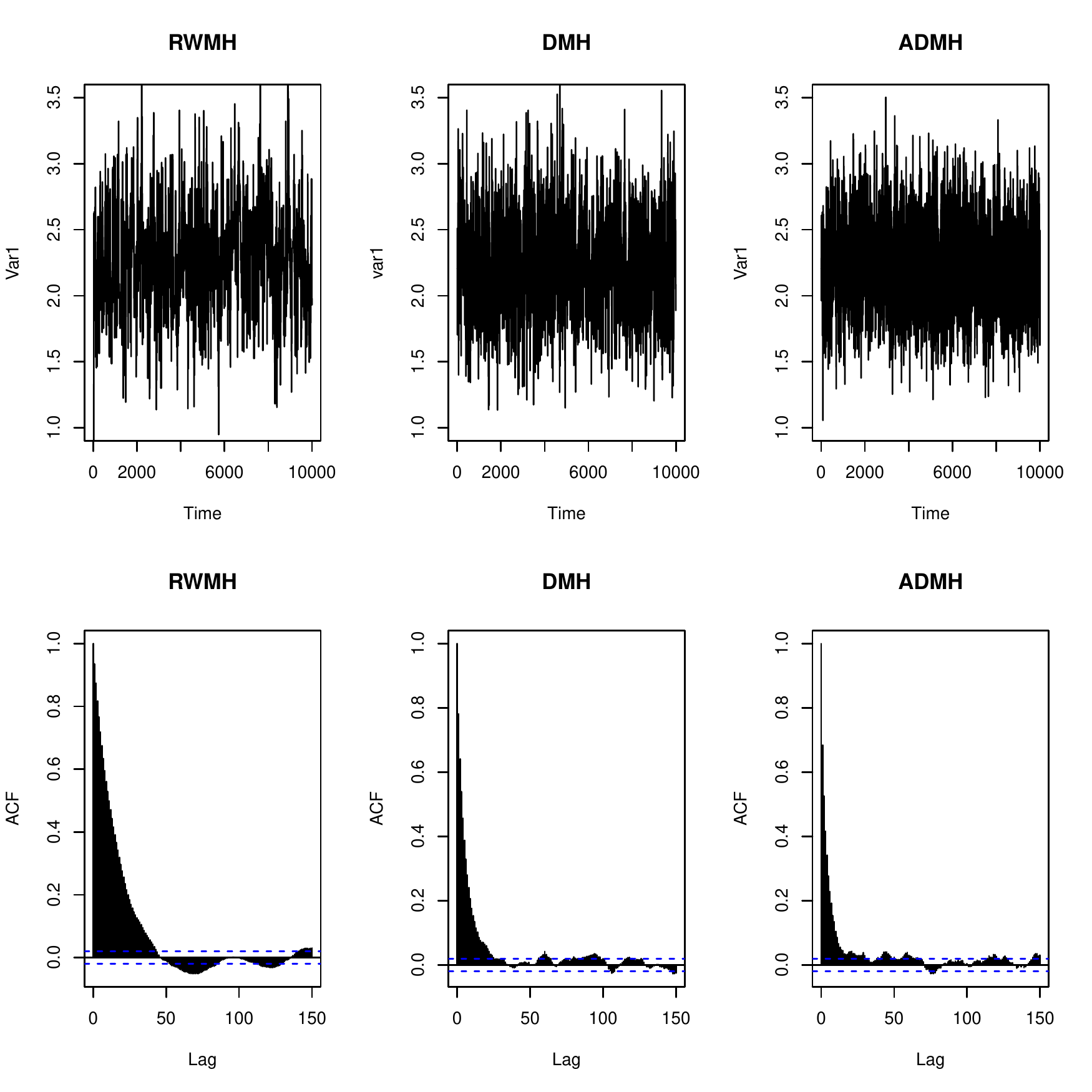}
  \caption{1a}
  \label{fig:sfig1}
\end{subfigure}%
\begin{subfigure}{.5\textwidth}
  \centering
  \includegraphics[width=\linewidth]{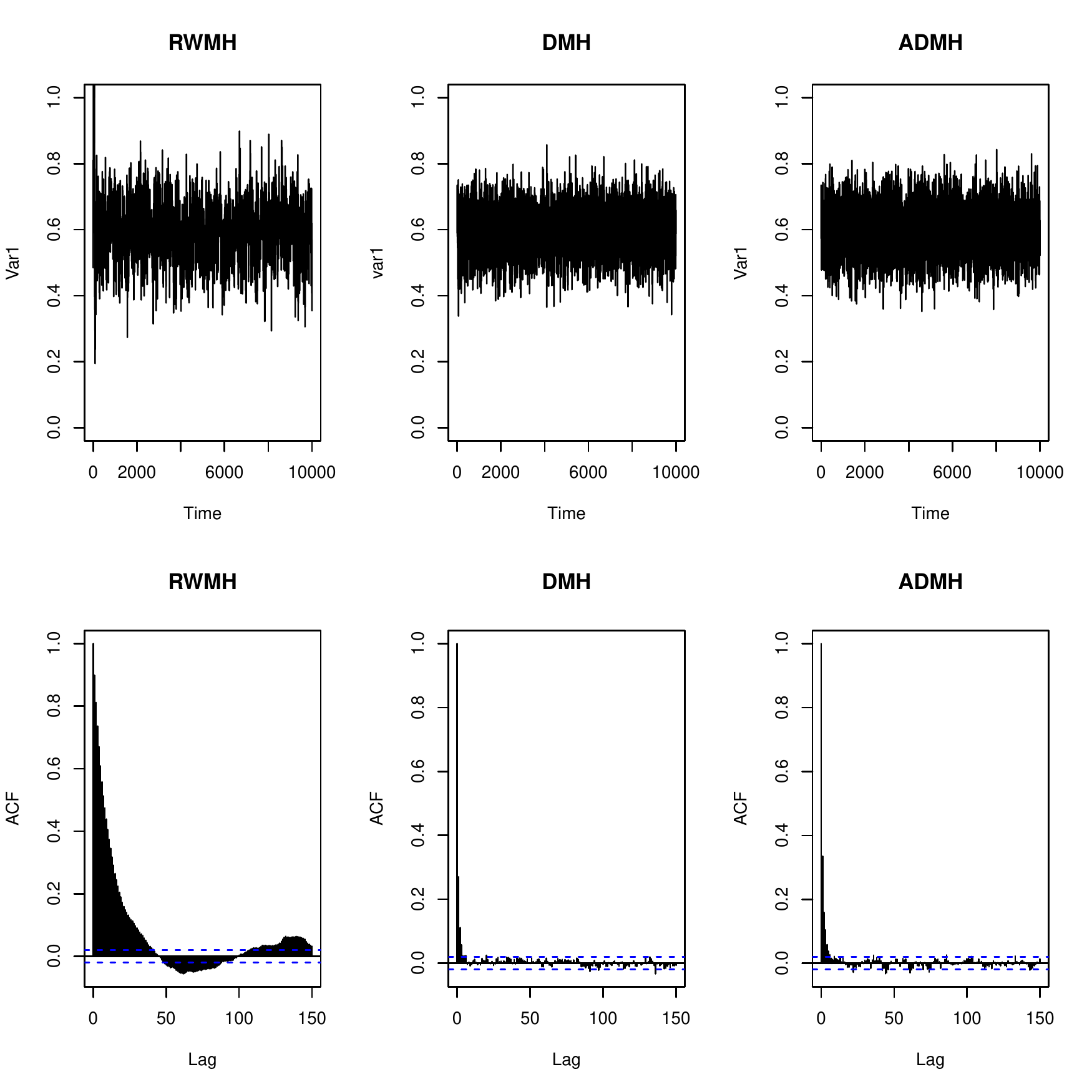}
  \caption{1b}
  \label{fig:sfig2}
\end{subfigure}
\caption{1a. Trace plot and ACF for Bernoulli model component 1. 1b. Trace plot and ACF for Normal model.}
\label{fig:fig}
\end{figure}

\begin{figure}
\begin{subfigure}{.5\textwidth}
  \centering
  \includegraphics[width=\linewidth]{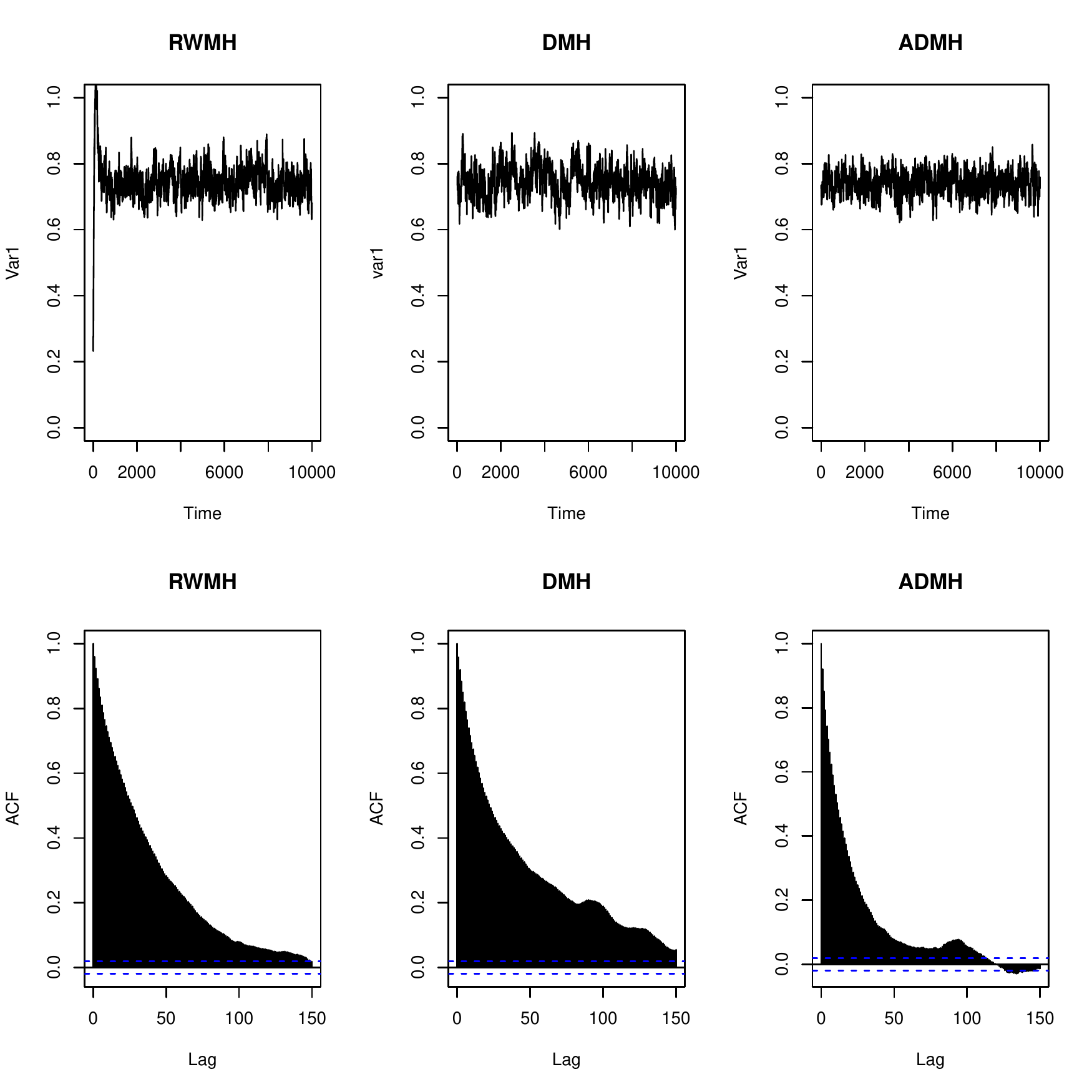}
  \caption{1c}
  \label{fig:sfig3}
\end{subfigure}%
\begin{subfigure}{.5\textwidth}
  \centering
  \includegraphics[width=\linewidth]{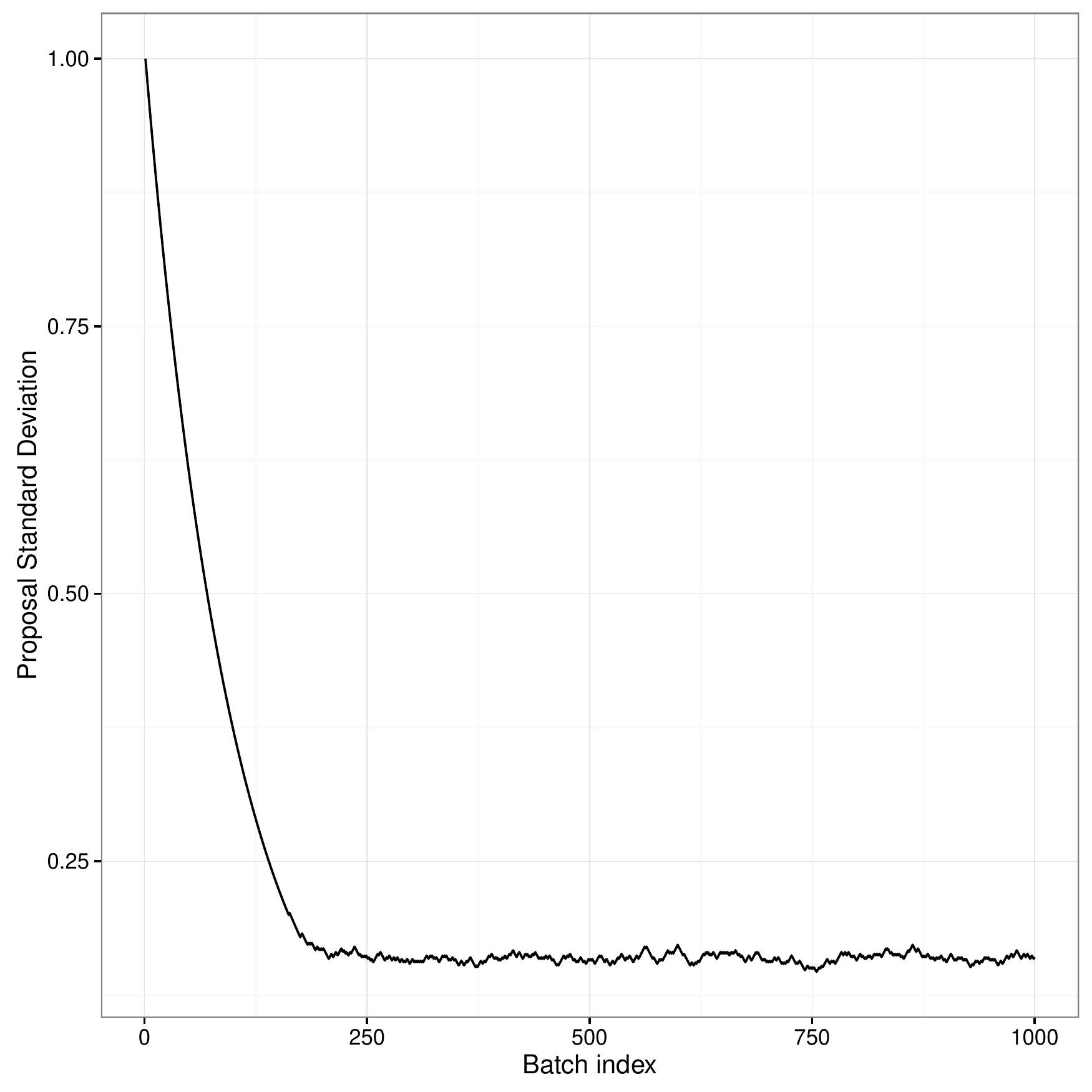}
  \caption{1d}
  \label{fig:sfig4}
\end{subfigure}
  \caption{1c. Trace plots and ACF for Poisson model. 1d. The proposal standard deviation plotted against the batch index.}
\label{fig:fig}
\end{figure}

\begin{proof}
	We need to check the two conditions given by \citet{roberts2007}. The \textit{Diminishing Adaptation} condition is satisfied as $\delta(b) \to 0$ as $b \to \infty$. The global maximal parameter value $M$ is specified beforehand. The proposal kernels are multivariate normals. So, for a large class of target densities which are log concave with bounded support, the \textit{Bounded Convergence} condition holds.
\end{proof}
The comparison of adaptive and non adaptive versions for ridge penalized regression model is presented in Table \ref{tab:esscomp} and \ref{tab:iactcomp}. We have started with a large value of the $\sigma$ parameter and ran the algorithm for a chain size of $10^5$ with a batch size of 100. We show how the adaption on $\sigma$ parameter stabilized with batch numbers in figure \ref{fig:sfig4}. In table \ref{tab:esscomp} we see that the adaptive version seems to have lower effective sample size and higher mean squared jumping distance than RWMH. Although, we started from a larger value of the $\sigma$ parameter, the parameter stabilized to a lower value, we have used a similar $\sigma$ in other two runs to make them comparable. In figure \ref{fig:sfig1}, \ref{fig:sfig2} and \ref{fig:sfig3} we are showing the autocorrelation function with lags, and it is evident that the DMH and Adaptive DMH both have sharper decreasing autocorrelation than RWMH.



	\section{Concluding Remarks} 
	\label{sec:conclusion}
	So, in conclusion, we have proposed a new kernel for Metropolis Hastings algorithm which changes its shape depending on the state. This class of algorithm has been considered in more detail in \citet{livingstone2015}, where he uses the term Position dependent proposal covariance. However, we consider one specific case of position dependence and explore the properties of that algorithm in detail. Specifically, we produce conditions of geometric ergodicity for this algorithm. Our algorithm is applicable whenever the derivative information of the target is available, and even in the case of its absence, it can be done numerically. In the case where simultaneous update is preferred, there our algorithm produces better mixing chains as demonstrated in our numerical studies. We also point out that this algorithm as a generalization of RWMH, provides more flexibility in tuning the sampler for various problems. We have included an adaptive version of the algorithm to automatize a part of it. Higher level of flexibility is always a better option to find the optimal configuration for a sampler. We provide implementations both the DMH and adaptive DMH algorithm in an R package.

\newpage
\bibliographystyle{plainnat}
\bibliography{references}

\begin{appendix}

\section{Proof of Lemma~\ref{lem:sigma}} \label{app:sigma lemma}
We have the fact that $\Lambda = diag\{s, 1, \cdots, 1\}$ and $G_d(x) = (g(x), g_1(x), \cdots, g_{d-1}(x)$. Let $G_{-1}(x)$ denote the $(d-1) \times (d-1)$ matrix of basis completion vectors $(g_1(x), \cdots, g_{d-1}(x))$. Then,
\begin{align*}
\Sigma(x) = G_d(x)\Lambda G_d(x)^T\\
=& (g(x), G_{-1}(x)) \Lambda (g(x), G_{-1}(x))^T\\
=& s g(x) g(x)^T + G_{-1}(x)G_{-1}(x)^T
\end{align*}

Now, we know that $G_d(x)$ is orthonormal, hence 

\begin{align*}
I_p = G_d(x)^T G_d(x) =& G_d(x)G_d(x)^T \\
 =& (g(x), G_{-1}(x))(g(x), G_{-1}(x))^T\\
 =& g(x) g(x)^T + G_{-1}(x)G_{-1}(x)^T
\end{align*}

Substituting $G_{-1}(x)G_{-1}(x)^T$ from above proves the lemma.

\section{Proof of Theorem~\ref{thm:ge}}\label{app:ge thm}
To prove Geometric Ergodicity, we follow the similar line of logic as Tweedie,1996 \citet{robe:twee:1996}. The following theorem for General state space Markov chains (Meyn and Tweedie, 1993)\citet{meyn:twee:2009} outlines our approach.

\begin{theorem}
	Suppose that $\Phi$ is a $phi$-irreducible and aperiodic. Then the following are equivalent:
	\begin{enumerate}[(i)]
		\item There is a function $V \geq 1$, finite for at least one $x$, and a small set $C$ such that, for some $\lambda_C < 1$, $b_C < \infty$, the drift condition: 
		\begin{align*}
		PV \leq \lambda_C V + b_C 1_C
		\end{align*}
		is satisfied, where $1_C$ denotes the indicator function of $C$. Here $PV(x) := \int V(y) P(x,dy)$.
		\item For some small set $C$ with $\phi(C) > 0$, there exists $\kappa > 1$, such that 
		\begin{align*}
		\sup_{x \in C} \mathcal{E}_x(\kappa^{\tau_C})
		\end{align*}
		\item The chain is Geometrically ergodic as there is a function $V \geq 1$, finite $f-$almost everywhere, which can be taken as the V in (i), and constants $\rho < 1$ and $R < \infty$ such that $V$ is finite, then,
		\begin{align*}
		||P^n(x,.) - f(.)||_{TV} \leq RV(x)\rho^n
		\end{align*}
	\end{enumerate}
	\label{thm:mnt}
\end{theorem}
\citet{robe:twee:1996} has shown that if $f(.)$ satisfies \ref{thm:mnt}, then Metropolis Hastings algorithm $P$ is Geometric Ergodic if and only if, there exists a real valued function $V > 1$, such that
\begin{align*}
\lim \sup_{||x|| \to \infty} PV(x)/V(x) < 1
\end{align*}

Our goal is to find a function $V(x)$ satisfying conditions mentioned in the above theorem. Using $V_{\tau}(x) = e^{\tau||x||}$. We have divided the state space into \emph{acceptance} region $A(x) = \{y: \alpha(x,y) = 1\}$ and \emph{possible rejection} regions $R(x) = A(x)^c$.

\begin{align*}
P V_{\tau}(x)/V_{\tau}(x) =& \int P(x,dy) V_{\tau}(y)/V_{\tau}(x)\\
=& \int_{A(x)} P(x,dy) V_{\tau}(y)/V_{\tau}(x) + \int_{R(x)} P(x,dy)\alpha(x,y) V_{\tau}(y)/V_{\tau}(x) \\
&+ \int_{R(x)} \delta_x(y) [1-\alpha(x,y)]P(x,dy)\\
\leq & \int_{\mathbb{R}^p} P(x,dy) V_{\tau}(y)/V_{\tau}(x) + \int_{R(x)} [1-\alpha(x,y)]P(x,dy)
\end{align*}

We can use the DMH kernel and expand the right hand side of the above equation. We use $||x||_s^2 := x^T S x$ notation to write the kernel. Here $\mu(x) = x + h \nabla \log f(x)$ denotes the center of the proposal at $x$. The inequality is sharpened by using a subset of the possible rejection region by intersecting with the interior of the point $x$.

\begin{align*}
\leq & (2\pi s)^{-p/2} \int_{\mathbb{R}^p} \exp\{-\frac{1}{2}||G_d(x)(y-\mu(x))||_s^2+\tau (||y|| - ||x||))\}dy \\
&+ (2\pi s)^{-p/2} \int_{R(x) \cap I(x)} \exp\{-\frac{1}{2}||G_d(x)(y-x)||_s^2\}dy\\
\end{align*}

Looking at the first term in the right hand side, we can use the fact that for $0 < s \leq 1$, $||G_x X||_s^2 \geq ||G_x X||^2 = ||X||^2$

\begin{align*}
& (2\pi s)^{-p/2} \int_{\mathbb{R}^p} \exp\{-\frac{1}{2}||G_d(x)(y-\mu(x))||_s^2+\tau (||y|| - ||x||))\}dy \\
\leq & (2\pi s)^{-p/2} \int_{\mathbb{R}^p} \exp\{-\frac{1}{2}||G_d(x)(y-\mu(x))||^2+\tau(||y|| - ||x||))\}dy \\
= & (2\pi s)^{-p/2} \int_{\mathbb{R}^p} \exp\{-\frac{1}{2}||(y-\mu(x))||^2+\tau(||y|| - ||x||))\}dy \\
\end{align*}

Multiplying the above term by $\exp(\tau(||x|| - ||\mu(x)||)$ and using triangle inequality, we get.

\begin{align*}
&= (2\pi s)^{-p/2} \int_{\mathbb{R}^p} \exp\{\tau(||x|| - ||\mu(x)||\}\exp\{-\frac{1}{2}[||(y-\mu(x))||^2-2t(||y|| - ||\mu(x)||)]\}dy \\
\leq & (2\pi s)^{-p/2} \int_{\mathbb{R}^p} \exp\{\tau(||x|| - ||\mu(x)||+2t^2\} \exp\{-\frac{1}{2}[||(y-\mu(x))||-\tau]^2\}dy\\
\end{align*}

Now $\limsup$ of the above term is less than 1. The second term converges to zero asymptotically, as $A(.)$ converges inwards in q. Hence,

\begin{align*}
\limsup_{||x|| \to \infty} PV(x)/V(x) < 1
\end{align*}

Provided that $\log f(x)$ is continuously differentiable, hence as it Chapter 6 of \citet{meyn:twee:2009} Meyn and Tweedie (1993), all compact sets are small, and thus it suffices to check the above condition for Geometric Ergodicity using Theorem 15.0.1 of \citet{meyn:twee:2009} Meyn and Tweedie (1993).

\section{Proof of Theorem~\ref{thm:rwtype}} 
\label{sec:proof_of_theorem_ref_thm_rwtype}
	The proposal density for DMH is 
	\begin{align*}
	P(X,.) = N_d(x+h\nabla \log f(x), tI_d+(s-1)G(x))
	\end{align*}
	We first use the following lemma similar to Theorem 2.2 of \citet{jarner2003}.
	\begin{lemma}
		Assume that $f(.)$ is a strictly positive and twice differentiable density, then for every $\epsilon > 0$, $\exists k > 0$, such that, 
		\begin{align*}
		P(x,B(x,k)) > 1 - \epsilon
		\end{align*}
		Where $B(x,k) := \{y \in \mathcal{X}: ||x-y|| < k\}$
		\label{lemma:rwlemma}
	\end{lemma}
	\begin{proof}
		As $f(.)$ is strictly positive and twice differentiable, $||\nabla \log f(x)||$ is bounded away from infinity. Hence the proposal is a normal distribution with finite mean and bounded marginal variances. So, for any $\epsilon$, we can always find a $k$ large enough so that $P(x,B(x,k)) > 1 - \epsilon$.
	\end{proof}
	Next, as the transition probability in $B(x,k)$ is given by 
	\begin{align*}
	P(x,B(x,k)) &= \int_{B(x,k)}[p(x,y)\alpha(x,y) + [1-\int_{\mathbb{R}^d} p(x,y)dy]]dy\\
				&\geq \int_{B(x,k)} p(x,y)\alpha(x,y) dy \\
				&= \int_{B(x,k)} p(x,y) \min\{\frac{p(y,x)f(y)}{p(x,y)f(x)},1\} dy \\
				&\geq \int_{B(x,k)} p(x,y) f(x) \min\{\frac{p(x,y) p(y,x)f(y)}{p(x,y)f(x)},p(x,y)\} dy \\
				&\geq \int_{B(x,k)} \min \{p(x,y)p(y,x)f(y),p(x,y)^2f(x)\} dy
	\end{align*}
	As $f(x)$ is positive in $\mathcal{X}$, there exists $\gamma$ such that $\inf_{y \in B(x,k)} \min (q(x,y),q(y,x)) > \gamma$. Hence the right hand side of the above equation is bounded below uniformly in $x$. So, for each $\epsilon > 0$, there exists a $k > 0$, such that $P(x,B(x,k)) > 1-\epsilon$, for all $x$. This result then follows from lemma~\ref{lemma:rwlemma}.


\end{appendix}

\end{document}